\documentclass[aps,prl,twocolumn,superscriptaddress,nofootinbib]{revtex4-1}
\usepackage{graphicx}
\usepackage{amssymb}
\usepackage[colorlinks=true,allcolors=black,pdfborder={0 0 0}]{hyperref}
\usepackage[all]{hypcap} 
\usepackage[usenames,dvipsnames,svgnames,table]{xcolor}

\begin{document}

\title{Adsorption-Induced Slip Inhibition for Polymer Melts on Ideal Substrates}

\author{Mark Ilton}
\affiliation{Department of Physics \& Astronomy, McMaster University, Hamilton, Ontario, Canada, L8S 4M1}
\affiliation{Polymer Science \& Engineering Department, University of Massachusetts Amherst, Amherst, MA, 01003, USA}

\author{Thomas Salez}
\affiliation{Univ. Bordeaux, CNRS, LOMA, UMR 5798, F-33405 Talence, France}
\affiliation{Global Station for Soft Matter, Global Institution for Collaborative Research and Education, Hokkaido University, Sapporo, Hokkaido 060-0808, Japan}
\affiliation{Laboratoire de Physico-Chimie Th\'{e}orique, UMR CNRS Gulliver 7083, ESPCI Paris, PSL Research University, 75005 Paris, France}

\author{Paul D. Fowler}	
\affiliation{Department of Physics \& Astronomy, McMaster University, Hamilton, Ontario, Canada, L8S 4M1}
\affiliation{Max Planck Institute for Dynamics and Self-Organization (MPIDS), Am Fa\ss berg 17,	37077 G\"{o}ttingen, Germany}

\author{Marco Rivetti}
\affiliation{Max Planck Institute for Dynamics and Self-Organization (MPIDS), Am Fa\ss berg 17,	37077 G\"{o}ttingen, Germany}

\author{Mohammed Aly}
\affiliation{D\'{e}partement de Physique, Ecole Normale Sup\'{e}rieure/PSL Research University, CNRS, 24 Rue Lhomond, 75005 Paris, France}

\author{Michael Benzaquen}
\affiliation{Laboratoire de Physico-Chimie Th\'{e}orique, UMR CNRS Gulliver 7083, ESPCI Paris, PSL Research University, 75005 Paris, France}
\affiliation{Ladhyx, UMR CNRS 7646, Ecole Polytechnique, 91128 Palaiseau Cedex, France}

\author{Joshua D. McGraw}
\affiliation{Department of Physics \& Astronomy, McMaster University, Hamilton, Ontario, Canada, L8S 4M1}
\affiliation{D\'{e}partement de Physique, Ecole Normale Sup\'{e}rieure/PSL Research University, CNRS, 24 Rue Lhomond, 75005 Paris, France}

\author{Elie Rapha\"{e}l}
\affiliation{Laboratoire de Physico-Chimie Th\'{e}orique, UMR CNRS Gulliver 7083, ESPCI Paris, PSL Research University, 75005 Paris, France}

\author{Kari Dalnoki-Veress}
\affiliation{Department of Physics \& Astronomy, McMaster University, Hamilton, Ontario, Canada, L8S 4M1}
\affiliation{Laboratoire de Physico-Chimie Th\'{e}orique, UMR CNRS Gulliver 7083, ESPCI Paris, PSL Research University, 75005 Paris, France}

\author{Oliver B\"aumchen}
\email{oliver.baeumchen@ds.mpg.de}
\affiliation{Max Planck Institute for Dynamics and Self-Organization (MPIDS), Am Fa\ss berg 17,	37077 G\"{o}ttingen, Germany}

\date{\today}
\begin{abstract}
Hydrodynamic slip of a liquid at a solid surface represents a fundamental phenomenon in fluid dynamics that governs liquid transport at small scales. For polymeric liquids, de Gennes predicted that the Navier boundary condition together with the theory of polymer dynamics imply extraordinarily large interfacial slip for entangled polymer melts on ideal surfaces; this Navier-de Gennes model was confirmed using dewetting experiments on ultra-smooth, low-energy substrates. Here, we use capillary leveling -- surface tension driven flow of films with initially non-uniform thickness -- of polymeric films on these same substrates. Measurement of the slip length from a robust one-parameter fit to a lubrication model is achieved. We show that at the lower shear rates involved in leveling experiments as compared to dewetting ones, the employed substrates can no longer be considered ideal. The data is instead consistent with physical adsorption of polymer chains at the solid/liquid interface. We extend the Navier-de Gennes description using one additional parameter, namely the density of physically adsorbed chains per unit surface. The resulting formulation is found to be in excellent agreement with the experimental observations. 
\end{abstract}     
\maketitle

When a liquid flows along a solid surface, molecular friction at the solid/liquid interface can have a large effect on the overall dynamics. For a sufficiently high solid/liquid interfacial friction, the fluid velocity parallel to the interface goes to zero at the boundary. This ``no-slip'' boundary condition is a  standard approximation for describing fluid flow at macroscopic length scales. In the past few decades, there have been many experiments measuring deviations from the no-slip boundary condition at microscopic length scales~\cite{Leger1999,Neto2005a,Lauga2007,Baumchen2010a,Rothstein2010}. These measurements have stimulated the interest in hydrodynamic slip for both the fundamental understanding of the molecular mechanisms involved, as well as the impact of slip on technological applications~\cite{Ajdari2006,Heryudono2007,Ren2008,Vermesh2009,Bottaro2014,Striolo2014,Ilton2015,Chen2015,Secchi2016,Inn1996,Pryamitsyn2006,Begam,Blake2006,Klein1994,klein1991}.

\begin{figure}[tp]
	\includegraphics[width=0.95\columnwidth]{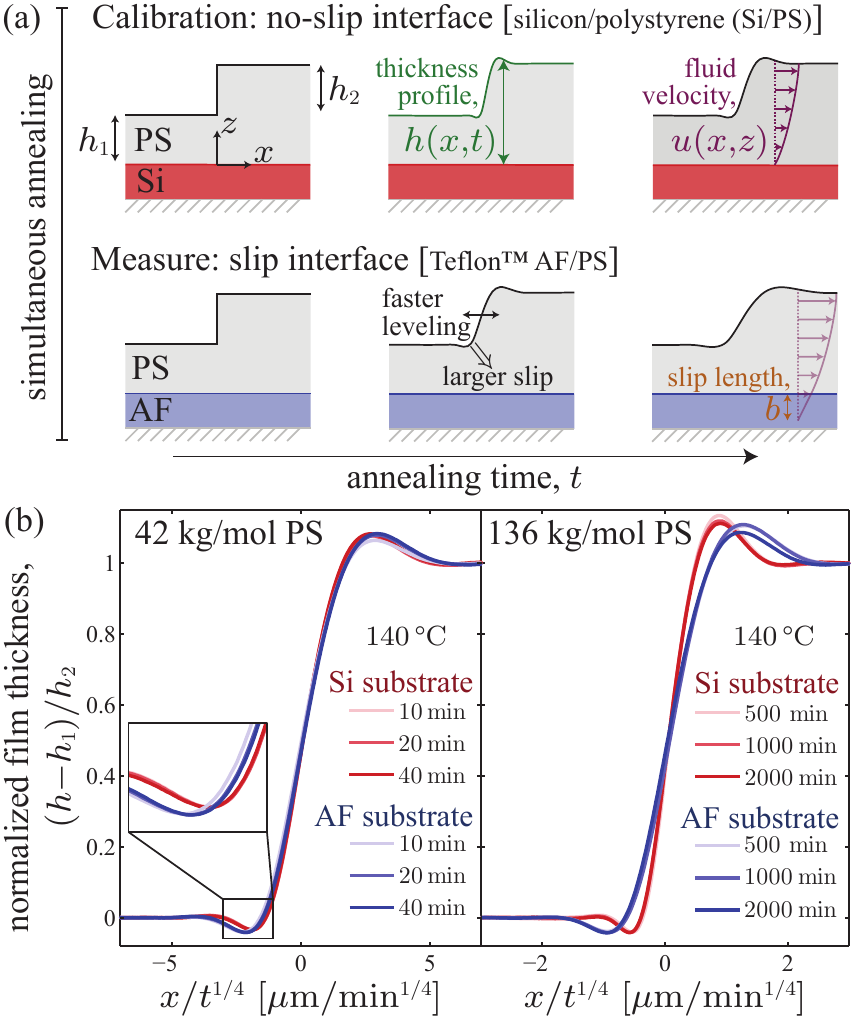}
	\caption{\label{fig:1} Interfacial slip causes a faster broadening of the film thickness profile in a capillary leveling experiment. (a) Schematic of the experimental procedure in which identical PS samples on two different substrates are annealed simultaneously. (b) Temporal series of experimentally measured atomic force microscopy profiles, normalized to demonstrate self-similarity of the film thickness profiles.}
\end{figure}

Hydrodynamic slip was first modelled by Navier~\cite{navierOLD} using a solid/liquid stress balance at the substrate, which can be used to define the slip length  $b =[u_x/\partial_z u_x]|_{z=0}$. As shown schematically in Fig.~\ref{fig:1}a, the linear extrapolation length of the horizontal fluid velocity profile to zero is the slip length, $b$. Experimental techniques used to quantitatively measure the slip length can be broadly classified into four main categories~\cite{Bouzigues2008}:  1) hydrodynamic drainage experiments, where the pressure is measured as fluid is squeezed out of (or drawn into) a small gap between two solid surfaces~\cite{Craig2001,Restagno2002,Vinogradova2003,Garcia2016},  2) a direct measurement of the velocity profile near the interface, by either using tracer particles~\cite{Ou2005,Joly2006,Li2015} or fluorescence recovery~\cite{Pit2000}, 3) pressure-drop experiments for flows driven through microchannels~\cite{cheng2002fluid,cheikh2003stick,choi2003apparent}, and 4) dewetting experiments, which measure the retraction of a thin layer of fluid from a low-energy substrate~\cite{Redon1994,Fetzer2005,Munch2011,Haefner2015}. 
Reported slip lengths from methods 1) to 4) vary from nm to mm~\cite{Granick2003,Wang1996}, while several studies also indicate the validity of the no-slip boundary condition ($b=0$)~\cite{Zettner2003,Honig2007,Schaeffel2013,Ahmad2015}. From these experiments, some of the parameters affecting the magnitude of hydrodynamic slip have been elucidated which include the interfacial properties~\cite{Barrat1999,Cho2004,Cottin-Bizonne2005}, surface roughness~\cite{Pit2000,Zhu2002b}, shear rate~\cite{Leger1999,mhetar1998slip,Craig2001}, and molecular weight in polymer fluids~\cite{Inn1996,Wang1996,Baumchen2009}. Thus, $b$ is highly sensitive to the solid/liquid combination. 

For polymer fluids flowing across ultra-smooth, low-energy surfaces, which act as ideal surfaces~\cite{DeGennes1979}, dewetting experiments have measured  large ($>1\mathrm{\ \mu m}$) slip lengths~\cite{Baumchen2009,Baumchen2012,Baumchen2014}. These results confirmed the scaling of the slip length with molecular weight for polymeric fluids, as originally predicted by de Gennes based on connecting the Navier boundary condition to polymer theory~\cite{DeGennes1979}. Specifically, Navier assumed that the substrate/liquid stress balance can be expressed as $\kappa u_x = \eta\partial_zu_x$, where $\kappa, u_x(x,z)$ and $\eta$ are the linear friction coefficient, the flow velocity parallel to the substrate and the viscosity, and $\partial_z$ denotes differentiation with respect to the surface-normal. De Gennes assumed a constant $\kappa$ and used the classic result of a strongly varying viscosity with molecular weight~\cite{rubinstein2003polymer}, $M_w$. These essential ingredients allowed the prediction of the \emph{Navier-de Gennes model} and subsequent experimental verification of the scaling law $b = \eta/\kappa\sim M_w^3$.

Here, we employ a different experimental technique with shear rates orders of magnitude smaller than those characteristic of dewetting and show that slip is inhibited even on an ideal surface, which is induced by adsorption of polymer chains. We use capillary leveling~\cite{buck2004decay,McGraw2012}, an experimental approach which invokes a film with an initially non-uniform thickness profile, resulting in an excess surface area compared to a flat film, the latter representing a metastable equilibrium. To suppress the energy cost of the excess surface area, surface tension causes the film to flow, driving it towards a uniform thickness. For an initially stepped film (see Fig.~\ref{fig:1}a), and after a transient regime, the surface profile evolves in a \emph{self-similar} fashion -- that is, flow causes the profile to broaden, but the characteristic shape remains fixed and the broadening is determined by a power law in time. By monitoring the self-similar profile and fitting it to a lubrication model, quantitative nano-rheological information about the film can be obtained. 

Previous works on capillary leveling measured the viscosity of thin polymer films supported by a substrate with a no-slip ($b=0$) boundary condition~\cite{McGraw2012}, and addressed the infinite-slip ($b\rightarrow\infty$) limit using freestanding polymer films where the two free interfaces provide no resistance to flow~\cite{Ilton2016}. In between the no-slip and infinite-slip extremes, the current work demonstrates the utility of capillary leveling as a quantitative probe of finite interfacial slip. We find that the measured slip length is independent of both temperature and sample geometry over the ranges studied. However, the slip length increases with the molecular weight of the polymer, and eventually saturates to a plateau at large molecular weight despite the ideal character of the substrate. This latter fact contrasts drastically with the previous high-shear-rate dewetting studies using identical materials and conditions~\cite{Baumchen2009,Baumchen2012,Baumchen2014,Haefner2015}. Inspired by the case of chemically-grafted substrates~\cite{Brochard-Wyart1996}, we propose a low-shear-rate description based on the presence of a dilute number of physically adsorbed polymer chains, which reconciles the two sets of experiments, and extends the Navier-de Gennes model described above. 

\section{Results}
\subsection{Experimental approach}

Capillary leveling of stepped films is used to measure the slip length of polystyrene (PS) on Teflon\texttrademark $\,$  fluoropolymer (AF) substrates (see Methods), a combination of materials which has been previously shown to exhibit interfacial slip~\cite{Baumchen2009,Baumchen2012,Baumchen2014,Haefner2015}. As a calibration, identical films of PS are prepared on silicon (Si) substrates since the Si/PS interface has no interfacial slip~\cite{McGraw2012}. Both types of samples are annealed simultaneously and side-by-side as outlined schematically in Fig.~\ref{fig:1}a, and surface profiles are obtained with atomic force microscopy (AFM). The self-similar profile of the Si/PS sample provides a calibration measurement of the PS capillary velocity, $v_c=\gamma/\eta$, where $\gamma$ and $ \eta$ are the surface tension and viscosity, respectively. Note that the value of $v_c$ depends on temperature and molecular weight, which are identical for the simultaneously studied Si/PS and AF/PS samples. The protocol thus allows the unambiguous and quantitative determination of the slip length of the solid/liquid (AF/PS) interface, the only differing quantity between the two simultaneously annealed samples. The measured film thickness profiles are self-similar in the reduced variable $x/t^{1/4}$, where $x$ is the horizontal coordinate and $t$ is the annealing time, for PS stepped films leveling on both substrates (Fig.~\ref{fig:1}b). We find that PS films broaden more rapidly on the AF substrates than on the Si calibration substrates (Fig.~\ref{fig:1}b) for all investigated molecular weights, and this faster broadening is more significant at higher PS molecular weight. In order to demonstrate that we can resolve even the smaller slip lengths, we show a zoom on the dip region of the lower molecular weight film as an inset. There, it can clearly be seen that the film on AF has also leveled further than that on Si. As it will be shown below (Fig.~\ref{fig:3}), we resolve slip lengths at the level of tens of nm. Capillary leveling thus provides one advantage over dewetting, for which small slip lengths have comparatively larger measurement error (Fig.~\ref{fig:3}).

\subsection{Theoretical approach}

To extract quantitatively the slip length at the solid/liquid interface, we employ a continuum hydrodynamic model for the thin liquid film. Using the incompressible Stokes' equations in the lubrication approximation~\cite{Oron1997}, and allowing for weak slip\footnote{We have also analyzed our experiments using the intermediate-slip thin-film equation outlined in Ref.~\cite{Munch2005}. In the worst case, this results in a small (on the order of $30\%$) increase in the measured slip length, which does not affect any of the conclusions of this work. Besides, we stress that strong-slip~\cite{Munch2005} or infinite-slip~\cite{Ilton2016} descriptions would be incompatible with the observed self-similarity (see Fig.~\ref{fig:1}b).} (slip length much smaller than the characteristic film thickness) at the solid/liquid interface, leads to a partial differential equation describing the evolution of the film thickness profile $h(x,t)$~\cite{Munch2005}:
\begin{equation}
\label{eq:WSTFE}
\frac{\partial h}{\partial t} = -\frac{v_c}{3}\frac{\partial}{\partial x}\left[ \left(h^3+3bh^2\right) \frac{\partial^3 h}{\partial x^3}\right].
\end{equation}

One can nondimensionalize this equation by introducing an arbitrary reference length scale $h_0\!=\!h_1+h_2/2$, and the associated time scale $3h_0/v_c$. Furthermore, for a given stepped initial profile (Fig.~\ref{fig:1}a), the rescaled solution $(h-h_1)/h_2$ of Eq.~(\ref{eq:WSTFE}) is self-similar in the variable~\cite{Munch2005,Salez2012a}:
\begin{equation}
U_0=\left(\frac{3x^4}{h_0^3v_ct}\right)^{1/4},
\end{equation}
but depends intrinsically on two parametric ratios, $h_2/h_1$ and $b/h_1$. As a particular case, for a stepped initial profile with $h_2/h_1\ll1$ one can linearize Eq.~(\ref{eq:WSTFE}). Nondimensionalizing the obtained equation by introducing the previous length scale $h_0$, but a different time scale $3h_0/[v_c(1+3b/h_0)]$, one obtains the result that the rescaled solution $(h-h_1)/h_2$ is now a single universal function of only the following generalized variable: 
\begin{equation}
U_b=\left[\frac{3x^4}{h_0^3v_c(1+3b/h_0)t}\right]^{1/4}.
\end{equation}

\begin{figure}[tp]
	\includegraphics[width=0.95\columnwidth]{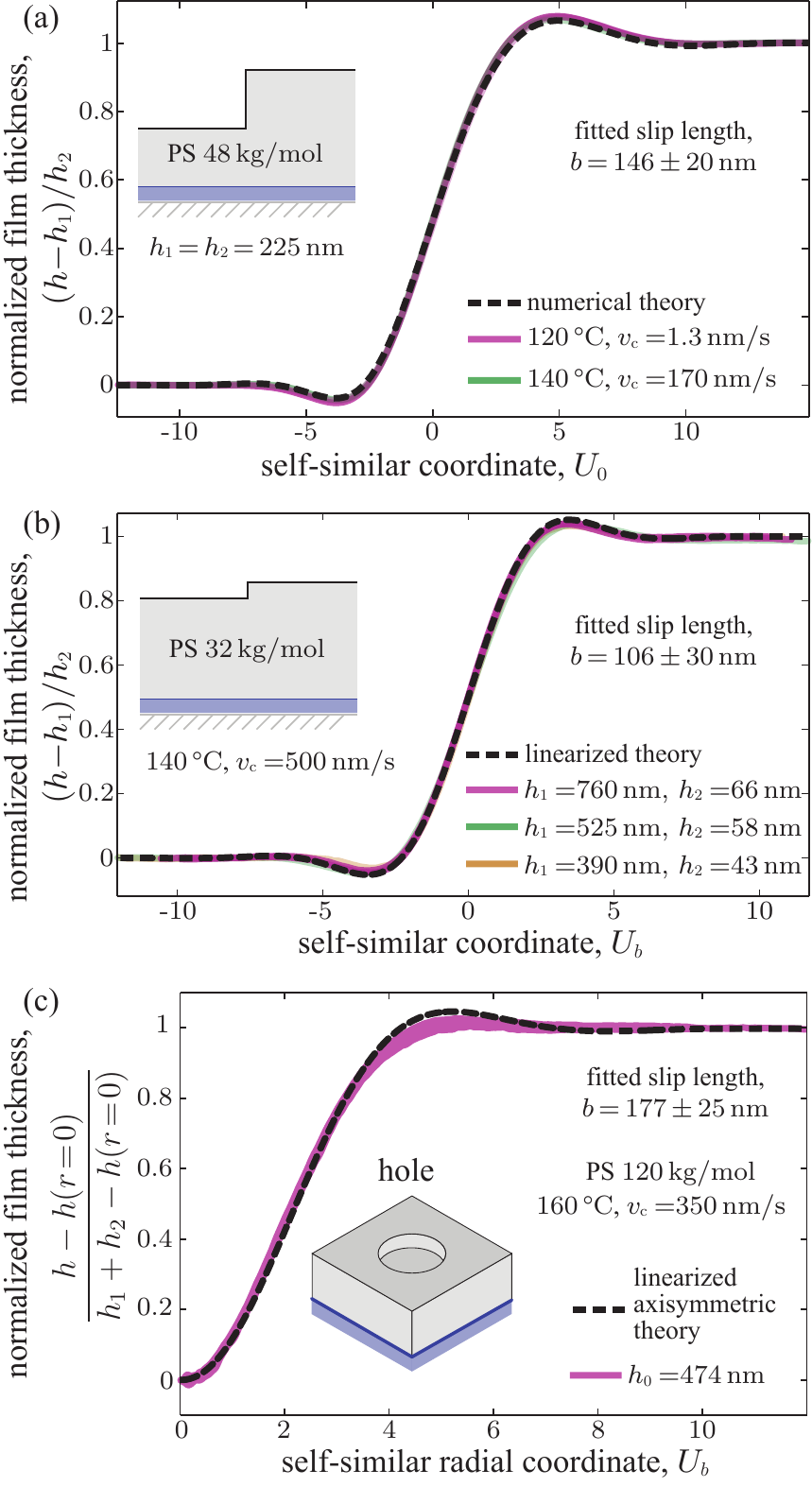}
	\caption{\label{fig:2} Capillary leveling is a robust experimental probe to measure slip length. For three different sample geometries, the rescaled self-similar theoretical profiles (dashed lines) fit the experimentally measured ones (solid lines) with one free parameter, the slip length $b$. Parameters are indicated in legends, and theoretical details are provided in main text. In (c), the position $x$ is replaced by the radial coordinate $r$.}
\end{figure}
In general, numerical solutions of Eq.~(\ref{eq:WSTFE})~\cite{Salez2012a} can be used to fit the data (Fig.~\ref{fig:2}a). For the particular case of $h_2\ll h_1$, analytical solutions of the linearized version of Eq.~(\ref{eq:WSTFE})~\cite{Salez2012d} can also be used to fit the data (Fig.~\ref{fig:2}b). Since $v_c$ is fixed by the simultaneous no-slip calibration experiment, and the sample geometry is directly measured using AFM, the slip length $b$ is the only free parameter in fitting the theory to experimentally measured AF/PS profiles (Fig.~\ref{fig:2}). The slip length is found to be independent of temperature (Fig.~\ref{fig:2}a) in the considered range, and is not sensitive to changes in the sample geometry through $h_1$ and $h_2$ (Fig.~\ref{fig:2}b). 

Complementary experiments were performed in a different geometry, in which the PS film was created with a cylindrical hole at the top~\cite{Backholm2014} (see Methods) instead of a step. The result is shown in Fig.~\ref{fig:2}c. The slip length is determined by fitting the radially averaged normalized profile to the analytical asymptotic solution of the linearized axisymmetric thin-film equation~\cite{Backholm2014} -- including weak slip through the variable $U_b$ above, where $x$ becomes the radial coordinate here.

\subsection{Effect of molecular weight on slip}

\begin{figure}[tp]
	\includegraphics[width=1\columnwidth]{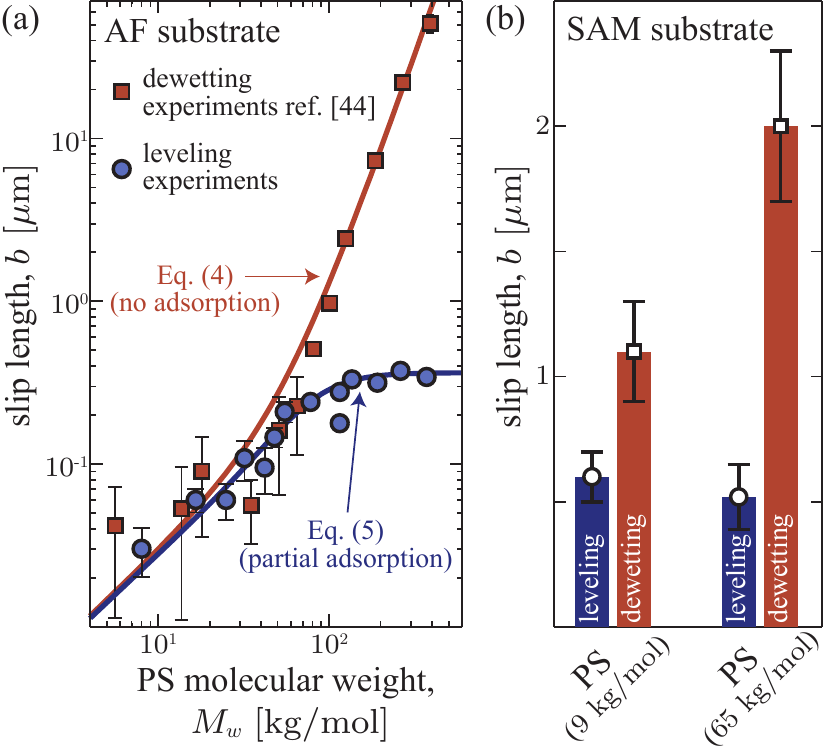}
	\caption{\label{fig:3} Slip on ideal substrates is inhibited at low shear rates due to adsorbing polymer chains. (a) Results from PS leveling experiments (blue circles) on AF substrates. Each data point consists of 2-12 individual measurements. For comparison, results from PS dewetting experiments (orange squares, data from~\cite{Baumchen2009}) on AF substrates are also shown. Two equations with one free parameter (solid lines) describe both sets of experiments: it assumes adsorption of chains in the low-shear-rate leveling experiments and no chain adsorption in the high-shear-rate dewetting experiments. (b) The difference in the measured slip length between leveling (blue) and dewetting (orange) experiments is confirmed using a different substrate (SAM) for PS (9\,kg/mol at $110\,^\circ$C and 65\,kg/mol at $135\,^\circ$C).}
\end{figure}

The effect of chain length on interfacial slip was studied using a series of 13 different PS molecular weights $8 \leq M_w \leq 373$\,kg/mol spanning the range between unentangled and well-entangled PS~\cite{rubinstein2003polymer, Fetters1999}. Results are shown in Fig.~\ref{fig:3}a (blue circles). At low molecular weight, the slip length increases with increasing PS molecular weight, but becomes approximately constant for molecular weights greater than $\sim100$\,kg/mol.

The molecular weight dependence of the AF/PS slip length has previously been found in dewetting studies (Fig.~\ref{fig:3}a, orange squares)  to increase sharply at large molecular weight~~\cite{Baumchen2009,Baumchen2012,Baumchen2014}, which contrasts with the leveling results (Fig.~\ref{fig:3}a, blue circles) in the current work. Although the results from the two techniques agree at low PS molecular weight, the leveling results exhibit slip lengths that are reduced by two orders of magnitude at the highest molecular weights. 

To determine if the difference in slip length at high molecular weight is specific to AF/PS, we performed a set of experiments with PS on self-assembled monolayer (SAM) substrates which are known to provide a slip boundary condition for PS~\cite{Fetzer2005,Baumchen2014}. In the SAM/PS experiments, both leveling and dewetting measurements were performed for two different molecular weights (9\,kg/mol and 65\,kg/mol). Results are shown in Fig.~\ref{fig:3}b. As for AF/PS, both molecular weights show a discrepancy between the slip length accessed with leveling and dewetting. Furthermore, the difference grows with molecular weight. Therefore, the observed difference in slip length between dewetting and leveling experiments exists also in the SAM/PS system, and is thus not specific to AF/PS.

\section{Discussion}

Using lubrication theory~\cite{Oron1997} applied to the leveling experiments, typical shear rates at the substrate can be estimated through $\partial_zu|_{z=0} =v_c h\partial_{x}^3h$. Using the experimental data (Fig.~\ref{fig:1}), we find strain rates of order $10^{-5}-10^{-6}\,\mathrm{s}^{-1}$ for the molecular weights used. This range is three orders of magnitude lower than the average shear rates calculated for dewetting with the same molecular weights~\cite{Baumchen2012}, and even lower if the maximum shear rate at the dewetting contact line is used. 

A quantitative analysis of the residence time for polymer molecules under flow illustrates the effects of the different shear rates between leveling and dewetting experiments (see Supplemental Information for details). The energy associated with the external force acting on an adsorbed polymer chain is estimated as $R_\mathrm{g}^{\,4}\partial_xP$, where $P$ denotes the (Laplace) pressure and $R_\mathrm{g}$ the radius of gyration of the chain. Typical values from the leveling experiments ($\partial_xP\approx 0.5\,$kPa/$\mu$m), with $R_\mathrm{g}$ = 29\,nm \cite{rubinstein2003polymer}, provide $R_\mathrm{g}^{\,4}\partial_xP\approx 3.5\times10^{-22}$\,J, which is substantially smaller than thermal energy $k_{\mathrm{B}}T\approx 5\times10^{-21}$J at $T = 150\,^\circ$C, or the van der Waals interaction energies. Conversely, the energy associated with dewetting experiments is larger than $k_{\mathrm{B}}T$ or van der Waals interaction energies by about one order of magnitude, thus crossing the threshold energy scale for chain desorption. The low shear rates for leveling thus lead to residence times for polymer chains at the solid/liquid interface that are long compared to the typical polymer relaxation time, and these adsorbed chains are expected to significantly contribute to the solid/liquid friction. This analysis is supported by studies which find a shear dependence of polymer adsorption~\cite{Cohen1982,McGlinn1988} and a recent work demonstrating that dewetting processes are faster when chain adsorption becomes weaker~\cite{Wang2017}. Assuming that such an adsorption scenario is operative in leveling but not in dewetting, the large difference in measured slip lengths at high molecular weights between the low-shear-rate leveling experiments and high-shear-rate dewetting experiments (Fig.~\ref{fig:3}) can be rationalized, as detailed below.

The Navier-de Gennes model~\cite{DeGennes1979} predicts that under ideal conditions of no adsorption, where the polymer melt slips along a smooth passive surface, the slip length follows the form:
\begin{equation}
\label{eq:b_ideal}
 b_\mathrm{ideal} = a \frac{\eta}{\eta_0} = a \frac{M_w}{M_0}\left[1+\left(\frac{M_w}{M_e}\right)^2\right], 
\end{equation}
where  $a$ is the monomer size, $\eta$ is the polymer-melt viscosity, $\eta_0$ is the viscosity of a melt of monomers, $M_0$ is the monomeric molecular weight, and $M_e$ is the entanglement molecular weight. The right-hand side of Eq.~(\ref{eq:b_ideal}) corresponds to a smooth interpolation between the Rouse and mean-field reptation regimes for the polymer-melt viscosity~\cite{rubinstein2003polymer}. To be consistent with the reference dewetting measurements~\cite{Baumchen2009,Henot2017}, we have chosen the  scaling of Eq.~(\ref{eq:b_ideal}) to match the large-molecular-weight limit ($b_\mathrm{ideal}\sim M_w^{\,3}$) found previously. We stress that the choice of a different power law (namely, the empirical reptation scaling $b_\mathrm{ideal}\sim M_w^{\,3.4}$) does not alter the conclusions of the present work. Using the parameters and data from Ref.~\cite{Baumchen2009} (\emph{i.e.} $a = 0.3$~nm, $M_0 = 104\,\mathrm{g/mol}$, $M_e/M_0 = 517$), we recall on Fig.~\ref{fig:3}a (orange line) that Eq.(\ref{eq:b_ideal}) agrees with the dewetting data over the entire molecular-weight range used. 

We now turn to the case of low-shear-rate experiments, and we describe the influence of transient physically adsorbed chains in an analogous fashion to the case of permanent chemically grafted chains~\cite{Brochard-Wyart1996}. In the dilute-adsorption regime, adding the adsorption-induced frictional stress of Eq.~(10b) from Ref.~\cite{Brochard-Wyart1996} to the previous ideal frictional stress $ \eta u|_{z=0}/b_\mathrm{ideal}$, within the Navier-de Gennes construction~\cite{DeGennes1979}, leads to the dilute-adsorption prediction for the slip length:
\begin{equation}
\label{eq:b_ads}
b_\mathrm{ads} =  \frac{b_\mathrm{ideal}}{1 + b_\mathrm{ideal}/b^\star}\ , 
\end{equation}
where $b^\star =  a M_e/(n M_0)$ and $n$ is the number of adsorbed chains in one cross-sectional chain area $\sim M_wa^2/M_0$. Invoking the parameters from Ref.~\cite{Baumchen2009} as above, the dimensionless number $n$ is thus the only unknown quantity, and we now make the assumption (justified from the fit below) that $n$ does not vary (or varies weakly) with $M_w$. Stated differently, the density of physically adsorbed chains per unit surface is assumed to scale inversely with the cross-sectional chain area.
 
By fitting Eq.~(\ref{eq:b_ads}) to the leveling experimental data in Fig.~\ref{fig:3}a, we find an excellent agreement (blue line) and extract
$n=0.45\pm0.02$. Therefore, with a single free parameter we are able to reconcile the two very different experimental measurements of the slip length on the same AF/PS system. The saturation value of the slip length at high molecular weights and for low shear rates appears to be set by $b^\star\sim N_e a$, where $N_e$ is the number of monomers between entanglements (omitting the numerical prefactor $1/n$ in front). The prefactor $1/n$ is expected to increase with shear rate, and to eventually diverge, thus allowing for a continuum of curves in between the two shown in Fig.~\ref{fig:3}a. In addition, the leveling data appears to have a sufficient resolution to observe for the first time the low-$M_w$ Rouse limit of the Navier-de Gennes prediction. We add two remarks: i) we self-consistently find $n<1$ which validates the dilute-adsorption hypothesis; ii) $n$ is indeed nearly constant, as having a variation of $n$ with $M_w$ would correspond to not having a plateau for $b$ at large $M_w$.

Although the substrates we use are very smooth (see Methods), it is reasonable to expect that chain adsorption may occur at least temporarily at low shear stresses. First, even ultra-smooth surfaces show contact-angle hysteresis: if a contact line can be pinned on atomic-scale roughness, then so too can polymer chains. Secondly, unfavorable wetting does not imply repulsive interaction between the solid and the liquid, as wetting is rather controlled by a balance between this interaction and the solid-air interaction. Finally, molecular dynamics simulations have shown that adsorbed groups of connected monomers can occur at unfavorable interfaces, and the length of these adsorbed groups increases with molecular weight~\cite{Smith2005}. Chains which are adsorbed for long enough durations of time to affect the interfacial fluid dynamics are likely to have multiple attached monomers. Therefore, larger adsorbed chains should exclude other chains from adsorbing to the substrate. The fact that $n$ is a constant smaller than 1 could be a signature of this exclusion mechanism. 

In conclusion, we have demonstrated that capillary leveling can quantitatively probe interfacial dynamics at low shear rates. The use of simultaneously-annealed measurement samples on AF substrates and calibration samples on no-slip Si substrates, combined with weak-slip lubrication theory, allow for a robust one-parameter-fit of the slip length to the experimental data. For the case of PS films on an AF substrate, we find the slip length to increase with PS molecular weight before reaching a plateau value at large molecular weights. This contrasts with previous dewetting measurements on the same AF/PS system, which showed a strong increase in slip length at large PS molecular weights, consistent with the Navier-de Gennes model. Inspired by previous results for grafted chains, we propose an extension of the Navier-de Gennes model which takes into account a dilute physical adsorption of polymer chains in the low-shear-rate leveling experiments, and no adsorption in the high-shear-rate dewetting experiments. With one free parameter, the proposed extension of the Navier-de Gennes model  is able to capture the molecular-weight dependence of the slip length for both sets of experiments. Beyond providing new fundamental insights on the actively-studied problem of hydrodynamic slip, these results demonstrate that even ultra-smooth low-energy surfaces such as Teflon cannot always be considered as ideal substrates.

\section{Methods}

\subsection{Substrate preparation and characterization}
Silicon (Si) wafers (obtained from University Wafer and Si-Mat) were cleaved into 1 cm $\times$ 1 cm squares. To create the calibration samples, the silicon wafers were  rinsed with ultra-pure water (18.2 M$\mathrm{\Omega}\,$cm, Pall), methanol, and toluene (Fisher Scientific, Optima grade). To create a slip substrate, the wafers were coated with a thin film of the amorphous fluoropolymer AF (AF1600/AF2400, obtained from Sigma Aldrich) by dip coating from a dilute solution (solvent FC-72, obtained from Acros Organics, 0.5\% w/w concentration solution, 0.5 $\,$mm/s retraction speed). Following the manufacturer's recommended procedure, the AF substrates were annealed for 20 minutes at $5\,^\circ\mathrm{C}$ above the glass-transition temperature of AF ($160\,^\circ\mathrm{C}$ for AF1600 or $240\,^\circ\mathrm{C}$ for AF2400) to remove residual solvent. The AF film thickness was 10-15$\,$nm, measured using ellipsometry (EP3, Accurion). Atomic force microscopy (AFM, Caliber, Veeco; Dimension and Multimode, Bruker) measurements showed that the AF substrates have a $0.3\,\mathrm{nm}$ RMS surface roughness, and that PS droplets have a Young's contact angle of $88^\circ$ on these substrates. As a second set of ultra-smooth, low-energy substrates, we decorated Si wafers with a dense self-assembled monolayer (SAM) of octadecyltrichlorosilane (OTS) and dodecyltrichlorosilane (DTS, both purchased from Sigma-Aldrich). The SAM was composed of a mixture of equal parts OTS and DTS, providing the largest slip length for low-molecular-weight PS, see~\cite{McGraw2017} for details. Silane molecules covalently bind to the native oxide layer of the Si wafer during the established procedure~\cite{Lessel2015, McGraw2017} for fabrication. These substrates have an RMS roughness of $0.2\,\mathrm{nm}$ and PS droplets have a long-time, receding contact angle of $63\,^\circ$ on these mixed OTS/DTS SAMs~\cite{McGraw2017}.

\subsection{Polymer film preparation}
Polystyrene (PS) with molecular weight ($M_w$) ranging between 8--373 kg/mol and polydispersity less than 1.1 was obtained from Polymer Source and PSS. Films with an initially stepped thickness profile (as in Fig.~\ref{fig:1}a) were made using a previously-described technique~\cite{McGraw2012}, with only minor modification. A bottom PS film (thickness range $h_1\!=\!100-800\,\mathrm{nm}$) and a top PS film (thickness range $h_2\!=\!40-400\,\mathrm{nm}$) were spun cast from a dilute toluene solution (liquid chromatography grade) onto freshly cleaved mica substrates (Ted Pella). The PS films were pre-annealed on mica in a home-built vacuum oven for at least ten times longer than the calculated longest relaxation time of the PS~\cite{rubinstein2003polymer} (pre-annealing temperature 140-180$\,^\circ\mathrm{C}$, pre-annealing time 4-72 hours; depending on the PS molecular weight). After annealing, the bottom PS film was floated onto an ultra-pure water bath (18.2 M$\mathrm{\Omega}\,$cm, Pall), and picked up onto either a silicon substrate (calibration) or AF substrate (measurement). The bottom film was then allowed to dry for at least 2 hours before undergoing another annealing (annealing for at least two times the calculated longest relaxation time) to relax residual stress.  The top PS film was then floated off its mica substrate onto the water bath. Sharp edges in the top PS film were created by the floating process for low $M_w$ PS~\cite{Baumchen2013}, or for high $M_w$ PS by a procedure which involved floating onto Si, cleaving, and refloating onto the water bath~\cite{McGraw2012}. The sharp-edged top film was then picked up off the water bath with the previously-prepared bottom PS film on a substrate. A final drying of the film at room temperature concluded the sample preparation procedure. Identical procedures were applied for the experiments on the SAM substrates.
Additional samples where the second film was prepared with a hole (as in Fig.~\ref{fig:2}c) were made in the same manner as the stepped films described above, except for the creation of sharp edges. For the hole geometry, a top film was floated onto the water bath and picked up using a metal washer with a millimetric circular hole to create a freestanding film. The top film was then heated above the PS glass-transition temperature in the freestanding state until small holes were nucleated with a diameter between $\sim 3$ and $10\,\mathrm{\mu m}$. After quenching to room temperature, the top film was transferred onto the bottom film supported by a solid substrate. Full details on the hole-geometry sample preparation are presented in Ref.~\cite{Backholm2014}. 

\subsection{Experimental setup}
Pairs of otherwise identically-prepared samples were used with only the substrate being different (Si or AF). The film thickness profile of each sample was determined by measuring the surface topography of the film using AFM, and averaging the 3D topography along the direction of translational quasi-invariance of the sample to obtain a 2D thickness profile. The pairs of samples were then placed side-by-side for simultaneous annealing in either the home-built vacuum oven or on a hot stage (Linkam, UK). For a given pair of samples, the annealing temperature was held constant (between 120 and 160$\,^\circ{}\mathrm{C}$), and chosen such that the PS was in its liquid melt state inducing the capillary-driven leveling of the thickness profiles. After a chosen duration of annealing $t$, the samples were rapidly quenched to room temperature, deep into the glassy state of PS, where the leveling process was temporarily halted. The broadening of the thickness profiles were measured using AFM. The samples were then further annealed, quenched, and measured again using AFM. The process of alternate annealing and AFM measurements was repeated until the measured thickness profiles became self-similar (Fig.~\ref{fig:1}b).

\section{Acknowledgements}
The authors thank Vincent Bertin, Pascal Damman, Fr\'ed\'eric Restagno, Barbara Wagner, Andreas M\"unch and Dirk Peschka for interesting discussions. They acknowledge financial support from the German Research Foundation (DFG) under grant BA3406/2 and NSERC (Canada). T.S. acknowledges financial support from the Global Station for Soft Matter, a project of Global Institution for Collaborative Research and Education at Hokkaido University. O.B. acknowledges financial support from the Joliot ESPCI Paris Chair and the Total-ESPCI Paris Chair. J.D.M. and M.A. were supported by LabEX ENS-ICFP: No. ANR-10- LABX-0010/ANR-10-IDEX-0001-02 PSL.

\section{Author Contributions}
M.I., T.S., J.D.M., E.R., K.D.-V. and O.B. conceived the project and designed research. M.I., P.D.F., M.R., M.A., J.D.M. and O.B. performed experiments. T.S., M.B., J.D.M. and E.R. developed the theory. All authors contributed to the analysis and interpretation of the results. M.I., T.S., J.D.M., E.R., K.D.-V. and O.B. wrote the manuscript.

\section{Additional information}

\noindent Supplementary Information accompanies this paper at [URL].\\

\noindent Competing financial interests: The authors declare no competing financial
interest.\\


%

\end{document}